# Voltage-controlled Cryogenic Boolean Logic Family Based on Ferroelectric SQUID

Shamiul Alam[1], Md Shafayat Hossain[2], Kai Ni[3], Vijaykrishnan Narayanan[4], and Ahmedullah Aziz[1*]

[1] Dept. of Electrical Eng. and Computer Sci., University of Tennessee, Knoxville, TN, 37996, USA
[2] Dept. of Physics, Princeton University, Princeton, NJ, 08544, USA
[3] Dept. of Electrical and Microelectronic Eng., Rochester Institute of Technology, Rochester, NY, 14623, USA
[4] School of Electrical Eng. and Computer Sci., Pennsylvania State University, University Park, PA, 16802, USA
[*]Corresponding Author. Email: aziz@utk.edu

***Abstract-* The recent progress in quantum computing and space exploration led to a surge in interest in cryogenic electronics. Superconducting devices such as Josephson junction, Josephson field effect transistor, cryotron, and superconducting quantum interference device (SQUID) are traditionally used to build cryogenic logic gates. However, due to the superconducting nature, gate-voltage-based control of these devices is extremely difficult. Even more challenging is to cascade the logic gates because most of these devices require current bias for their operation. Therefore, these devices are not as convenient as the semiconducting transistors to design logic gates. Here, to overcome these challenges, we propose a ferroelectric SQUID (FeSQUID) based voltage-controlled logic gates. FeSQUID exhibits two different critical current levels for two different voltage-switchable polarization states of the ferroelectric. We utilize the polarization-dependent (hence, voltage-controllable) superconducting to resistive switching of FeSQUID to design Boolean logic gates such as Copy, NOT, AND, and OR gates. The operations of these gates are verified using a Verilog-A-based compact model of FeSQUID. Finally, to demonstrate the fanning out capability of FeSQUID-based logic family, we simulate a 2-input XOR gate using FeSQUID-based NOT, AND, and OR gates. Together with the ongoing progress on FeSQUID-based non-volatile memory, our designed FeSQUID-based logic family will enable all- FeSQUID based cryogenic computer, ensure minimum mismatch between logic and memory blocks in terms of speed, power consumption, and fabrication process.**

The idea of superconducting electronics (SCE) was first put forward in 1950s with the effort of building magnetic field-modulated superconducting wires [1,2]. The requirement of extremely low temperature was one of the major roadblocks for the superconducting devices to be used in practical applications which now becomes an advantage, thanks to the recent interests in superconducting qubit-based quantum computers and space electronics [3–6]. Superconducting logic circuits can significantly improve the performance of the control processor in the quantum computers and the aircrafts for space exploration [3]. Even in the classical computing, SCE shows immense potential. Although CMOS technology can now allow us to put more than trillion transistors on a single chip, the power dissipation reaches the physical limit and makes further scaling difficult [6,7]. To solve this issue, several beyond-CMOS alternatives are being explored. SCE is considered as one of the most promising alternatives to CMOS technology, thanks to their high speed and low power operation [2,8]. SCE has also been used in radio frequency receivers, high-end computing, and so on [9,10].

Over the last few decades, several superconducting logic devices [1,11–17] have been introduced. All of these except Josephson junction (JJ) have been limited within the basic characterization. JJ has been successfully utilized in a wide variety of applications due to their high speed (hundreds of gigahertz) and low power consumption (sub-aJ/bit). However, JJ-based circuits suffer from challenges: (i) the manipulation of single flux quanta causes issues while driving large impedances or fanning out digital signals, (ii) difficulty in the fabrication of uniform JJs over a wafer and the effects of magnetic field limit the distance between two JJs and hence, the integration density [18]. Another disadvantage (particular to logic circuits) of two-terminal JJ and superconducting quantum interference device (SQUID) is that the absence of resistance in superconductors hinders the voltage biasing and hence, superconducting devices cannot offer convenient gating mechanism like the semiconducting transistors. Therefore, there has been a significant amount of effort to implement a gate-tunable superconducting device [1,12–14,17].

SQUID is one of the basic building blocks for superconducting circuits and systems. Researchers have tried continuously to introduce gate-tunability to SQUIDs. The efforts include utilizing ionic liquid [19,20], integrating ferromagnetic components [21,22], and so on. Although the integration of ferromagnetic components allows non-volatile tunability, magnetic tunability limits the scalability of the circuits and systems. Recently, a new technique has been demonstrated that uses a ferroelectric material to tune the superconductivity of the SQUID [23]. Ferroelectric materials show voltage-controlled non-volatile switching of the polarization. Therefore, the incorporation of the ferroelectric material introduces voltage-controlled non-volatile switching of the critical current of the SQUID which opens a lot of possibility of utilizing this device in circuit/system level. This ferroelectric SQUID (FeSQUID) has already been used to design a cryogenic memory system, which promises voltage-controllability, non-volatility, scalability, separate read write paths, and energy efficient operation [24]. Therefore, FeSQUID-based cryogenic memory might be able to solve the issues of existing cryogenic memories and pave the way of large-scale development of quantum computers and SCE [3,6,25–28]. In this work, we present a voltage-controllable cryogenic Boolean logic family based on FeSQUIDs that does not impose any limit on the fan-out. FeSQUID-based logic family will allow the close integration of logic and memory blocks since both logic and memory can be designed using the same device. Therefore, there will be no mismatch between the logic and memory blocks in terms of speed, power consumption, and fabrication process which will also reduce the 'memory wall' bottleneck [29–31].

We start our discussion with the overview of FeSQUID. A SQUID built with two weak links in parallel is fabricated on top of a ferroelectric material [schematic shown in Fig. 1(a)]. The choice of superconducting and ferroelectric materials depends on lattice matching between them [32,33]. Ref. [23] demonstrated a FeSQUID device by fabricating a 15 nm thick $\alpha Mo_{80}Si_{20}$ planar SQUID and 70 nm thick $PbZr_{0.2}Ti_{0.8}O_3$ (PZT) interface on a 15 nm thick $SrRuO_3$ bottom electrode. Ferroelectrics are dielectric materials that can retain polarization switchable by external voltage/electric field. Fig. 1(b) shows the voltage-controlled non-volatile switching of the $P_{FE}$ of a PZT ferroelectric. In a ferroelectric material, internal $P_{FE}$ translates into a surface charge that induces an electric field and injects direct charge [34]. Now, when a SQUID is fabricated on top of a ferroelectric, the superconducting layer screens the charge bound in the interface. This bound charge in the surface of the interface ($\oiint P_R . dA$, where $P_R$ is the remnant $P_{FE}$ and $A$ is the surface area) directly depends on the remnant $P_{FE}$ – negative $P_R$ ($P_R^-$) increases the bound charge in the surface of the interface whereas positive $P_R$ ($P_R^+$) results in an opposite effect [23,32]. The resulting change in the carrier density impacts the critical temperature ($T_C$) and hence, the superconducting energy gap ($\Delta$). The dependence of $\Delta$ on $T_C$ is given by Bardeen–Cooper–Schrieffer (BCS) theory [35,36]-

$$\Delta\,(T) = 1.763 k_B T_C\, tanh\left(2.2\sqrt{\frac{T_C}{T} - 1}\right), \quad (1)$$

where $T$ is the temperature and $k_B$ is the Boltzmann constant. According to the Ambegaokar–Baratoff (AB) theory [37], $\Delta(T)$ determines the critical current ($I_C$) as given by-

$$I_C = \frac{\pi\Delta}{2q_e R_N}\tanh\left(\frac{\Delta}{2k_B T}\right), \quad (2)$$

where $q_e$ is the charge of an electron and $R_N$ is the normal state resistance of the SQUID. Therefore, based on the two non-volatile polarization states of the ferroelectric (controlled by voltage), two levels of $I_C$ are observed in the $I$-$V$ characteristics of the SQUID- $P_R^-$ and $P_R^+$ states of the ferroelectric lead to high ($I_{C,high}$) and low ($I_{C,low}$) $I_C$ for the SQUID, respectively [Fig. 1(c)]. In this work, we use this voltage-controllability of SQUID's superconductivity to design the basic Boolean logic gates for cryogenic applications. Fig. 1(d) illustrates how we utilize the FeSQUID to design logic gates. Here, we use the polarization states ($P_R^-$ and $P_R^+$) of the ferroelectric to define the logic states (logic '0' and '1', respectively) which can be switched by applying suitable voltage input. Although the coercive voltage of PZT is ~3 V [Fig. 1(b)], in our design, we use much larger voltage as the input (−6 V or +6 V for logic '0' and '1', respectively) to ensure that the ferroelectric remains in either $P_R^-$ or $P_R^+$. These voltage levels can be reduced using different ferroelectric material with scaled dimensions. In our design, logic '0' and '1' correspond to $I_{C,high}$ and $I_{C,low}$, respectively. Now, if we bias the FeSQUID with a bias current ($I_{C,low} < I < I_{C,high}$), FeSQUID with logic '0' ('1') input corresponds to superconducting (non-superconducting) state of the SQUID. Based on the state of the SQUID, the bias current will flow either through the SQUID or through the external circuitry. We harness this phenomenon to design the FeSQUID-based logic gates (discussed later in details).

In this work, we demonstrate the operation of our designed FeSQUID-based logic gates through simulation. For this, we first develop a circuit compatible compact model for FeSQUID in Verilog-A and calibrate that with the experimental observation. We first model the non-volatile switching of ferroelectric polarization using the following Preisach model equation-

$$P_{FE} = P_S \tanh\left[\frac{1}{2E_C}\ln\left(\frac{P_S+P_R}{P_S-P_R}\right)(E_{FE} \mp E_C) + P_{off}\right], \quad (3)$$

where $P_S$, $P_R$, and $P_{off}$ are the saturation, remnant, and offset polarizations, respectively. $E_C$ and $E_{FE}$ represent the coercive and applied electric fields across the ferroelectric, respectively. Figure 1(b) shows the validation of the $P_{FE}$-$V$ characteristics of PZT with the experimental result reported in Ref. [23]. Next, we phenomenologically (look-up-table-based approach) implement the dependence of $T_C$ on $P_{FE}$. Then we use equation (2) to calculate Δ(T) and subsequently equation (3) to calculate $I_C$ of the SQUID. Finally, we model the $I$-$V$ characteristics of the SQUID by utilizing the following equation-

$$V = \begin{cases} 0, & I < I_C \\ I \times R_N, & I \geq I_C \end{cases} \quad (4)$$

For any logic gate, a crucial requirement is the capability of handling fan-out. For a cascadable logic system, the outputs of the logic gates should be able to drive the inputs of the next stage. Now, for FeSQUID, to drive one gate by the output of another gate, the output voltage needs to be sufficiently large to set the intended polarization in the ferroelectric of the FeSQUID. Figure 2(a) shows the schematic of a FeSQUID-based copy gate. As seen in Fig. 1(c), the $I_C$ values of the FeSQUID device are 2.6 $\mu A$ and 4 $\mu A$, and the values of $R_N$ are $0.95\ k\Omega$ and $1.75\ k\Omega$ for $P_R^-$ and $P_R^+$, respectively. For proper operation of the copy gate shown in Fig. 2(a), the value of the external resistance ($R$) needs to be chosen in a way so that when the FeSQUID becomes resistive, almost all the bias currents flow through $R$. Also, we have to choose a bias current so that it satisfies $I_{C,low} < I < I_{C,high}$. Here, we choose $3.2\ \mu A$ and $10\ \Omega$ for $I$ and $R$, respectively. For these chosen values, we get $\sim 31\ \mu V$ (0 V) for logic '1' ('0') at the output of the copy gate shown in Fig. 2(a) which are not sufficient to drive the FeSQUID of the next stage. To circumvent this issue, we utilize heater cryotron ($hTron$) [18,28,38] at the output of each logic gates to develop a cascadable logic system based on FeSQUIDs. Figure 2(c) displays the I-V characteristics of $hTron$. $hTron$ is a three terminal gate current controlled superconducting switch, consisting of a superconducting channel and a resistive gate. The $hTron$ channel initially remains superconducting but a high enough (larger than the switching current) gate current can switch the superconducting channel to the resistive state. In our simulation, we utilize a phenomenological compact model for $hTron$ [28]. After using $hTron$ (working principle of the designed logic gates are discussed later), the output voltage of the copy gate becomes $-6\ V$ and $+6\ V$ for logic '0' and '1', respectively as required to drive the FeSQUID of the next stage [Fig. 2(b)].

It is worthwhile to discuss the design methodology, working principle, and simulated results of FeSQUID-based logic family. First, we design 1-input Copy gate utilizing one FeSQUID where the input is applied as voltage across the ferroelectric. Figure 3(a) shows the schematic of the Copy gate. In case of logic '0' ($-6$ V) applied to the input and suitable bias current ($I$) is applied, the FeSQUID shows $I_{C,high}$ ($>I$) and becomes superconducting [Fig. 3(b)]. Therefore, almost all the bias current flows through FeSQUID and the $hTron$ gate does not get enough current to switch the channel to its resistive state [Fig. 4(b)]. Due to the proper biasing of $hTron$, we get logic '0' ($-6$ V) at the output. Now, for logic '1' (+6 V) at the input, FeSQUID becomes resistive, driving most of the bias current to the gate of $hTron$ which switches the channel to resistive state [Fig. 3(c)], and hence, we obtain logic '1' (+6 V) at the output.

Next, we design the NOT gate whose schematic is presented in Fig. 3(d). In the NOT gate, the gate of the $hTron$ is connected in series with the FeSQUID. Therefore, for logic '0' ($-6$ V), FeSQUID becomes superconducting and passes enough current to the gate of $hTron$ which consequently switches the channel to its resistive state and we get logic '1' (+6 V) at the output [Fig. 3(e)]. For logic '1' (+6 V) at the input, the exact opposite scenario occurs for the FeSQUID and $hTron$, and we get logic '0' ($-6$ V) [Fig. 3(f)]. Figure 3(g) presents the simulation results for the 1-input Copy and NOT gates.

Having designed the 1-input gates, next we design the 2-input AND and OR gates where we use two FeSQUIDs and apply the inputs as voltages across the ferroelectric materials of the FeSQUIDs. Figure 4(a) captures the schematic of the AND gate where we connect two FeSQUIDs in parallel and hence, if any of the FeSQUIDs become superconducting ('00', '01', and '10' cases), most of the bias current flows through that and deprives the $hTron$ gate to get enough current to switch the channel to resistive [Figs. 4(a) & (b)]. Consequently, we get logic '0' ($-6$ V) at the output. However, for the '11' case at the input, both the FeSQUIDs switch to the resistive state driving the bias current through the $hTron$ gate which creates logic '1' (+6 V) at the output [Fig. 4(c)]. On the contrary, in the OR gate, two FeSQUIDs are connected in series [Fig. 4(d)]. Therefore, only when both the FeSQUIDs remain superconducting ('00' case), the $hTron$ gate does not experience enough current to switch the superconducting channel and creates an output of logic '0' ($-6$ V) [Fig. 4(d)]. For other cases ('01', '10', and '11'), at least one FeSQUID becomes resistive and hence, the $hTron$ channel switches to the resistive state and we get logic '1' (+6 V) at the output [Figs. 4(e) & (f)]. Lastly, to demonstrate the capability of FeSQUID-based logic family to handle fan-out, we simulate a 2-input XOR gate using the FeSQUID-based NOT, AND, and OR gates. Schematic of the XOR gate is shown in Fig. 4(g). Figure 4(h) shows the simulation results for the 2-input AND, OR, and XOR gates.

In summary, using voltage-controlled superconductivity in FeSQUIDs, we developed CMOS-like voltage-controlled superconducting logic family (1-input Copy and NOT gates, and 2-input AND and OR gates), which has immediate potential applications in superconducting electronics, quantum computers, and space exploration. Moreover, our designed logic circuits are transferrable to FeSQUIDs with different ferroelectric and superconducting materials as well as other voltage-controlled superconducting devices such as Josephson junction FET [39], Dayem bridge transistors [17,40], etc.

**Acknowledgement**

The authors thank Michael L. Schneider and Matthew R. Pufall of NIST for the helpful scholarly discussions.

**Data Availability**

The data that support the plots within this paper and other finding of this study are available from the corresponding author upon reasonable request.

**Author Contributions**

S.A. conceived the idea, designed the logic gates, and performed the simulations. M.S.H., K.N., V.N., and A.A. analyzed and helped finalizing the designs. A.A. supervised the project. All authors commented on the results and wrote the manuscript.

**Competing Interests**

The authors declare no competing interests.

**References**

1. Buck, D. A. *The Cryotron—A Superconductive Computer Component*. *Proc. IRE* **44**, 482–493 (1956). doi:10.1109/JRPROC.1956.274927.
2. Likharev, K. K. *Superconductor digital electronics*. *Phys. C Supercond. its Appl.* **482**, 6–18 (North-Holland, 2012). doi:10.1016/J.PHYSC.2012.05.016.
3. Tannu, S. S., Carmean, D. M. & Qureshi, M. K. Cryogenic-DRAM based memory system for scalable quantum computers: A feasibility study. in *ACM Int. Conf. Proceeding Ser.* (2017). doi:10.1145/3132402.3132436.
4. Dicarlo, L., Chow, J. M., Gambetta, J. M., Bishop, L. S., Johnson, B. R., Schuster, D. I., Majer, J., Blais, A., Frunzio, L., Girvin, S. M. & Schoelkopf, R. J. *Demonstration of two-qubit algorithms with a superconducting quantum processor*. *Nature* (2009). doi:10.1038/nature08121.
5. Kessler, M. F. *The Infrared Space Observatory (ISO) mission*. *Adv. Sp. Res.* (2002). doi:10.1016/S0273-1177(02)00557-4.
6. Alam, S., Hossain, M. S., Srinivasa, S. R. & Aziz, A. *Cryogenic Memory Technologies*. *ArXiV Prepr.* (2021). doi:10.48550/arXiv.2111.09436.
7. Masanet, E., Shehabi, A., Lei, N., Smith, S. & Koomey, J. *Recalibrating global data center energy-use estimates*. *Science (80-. ).* **367**, 984–986 (2020). doi:10.1126/science.aba3758.
8. Ayala, C. L., Tanaka, T., Saito, R., Nozoe, M., Takeuchi, N. & Yoshikawa, N. *MANA: A Monolithic Adiabatic iNtegration Architecture Microprocessor Using 1.4-zJ/op Unshunted Superconductor Josephson Junction Devices*. *IEEE J. Solid-State Circuits* **56**, 1152–1165 (Institute of Electrical and Electronics Engineers Inc., 2021). doi:10.1109/JSSC.2020.3041338.
9. Mukhanov, O. A., Kirichenko, D., Vernik, I. V., Filippov, T. V., Kirichenko, A., Webber, R., Dotsenko, V., Talalaevskii, A., Tang, J. C., Sahu, A., Shevchenko, P., Miller, R., Kaplan, S. B., Sarwana, S. & Gupta, D. *Superconductor digital-RF receiver systems*. *IEICE Trans. Electron.* (2008). doi:10.1093/ietele/e91-c.3.306.
10. Vernik, I. V., Kirichenko, D. E., Dotsenko, V. V., Miller, R., Webber, R. J., Shevchenko, P., Talalaevskii, A., Gupta, D. & Mukhanov, O. A. Cryocooled wideband digital channelizing radio-frequency receiver based on low-pass ADC. in *Supercond. Sci. Technol.* (2007). doi:10.1088/0953-2048/20/11/S05.
11. Josephson, B. D. *Possible new effects in superconductive tunnelling*. *Phys. Lett.* (1962). doi:10.1016/0031-9163(62)91369-0.
12. Lee, S. B., Hutchinson, G. D., Williams, D. A., Hasko, D. G. & Ahmed, H. *Superconducting nanotransistor based digital logic gates*. *Nanotechnology* **14**, 188 (IOP Publishing, 2003). doi:10.1088/0957-4484/14/2/317.
13. Akazaki, T., Takayanagi, H., Nitta, J. & Enoki, T. *A Josephson field effect transistor using an InAs-inserted-channel In0.52Al0.48As/In0.53Ga0.47As inverted modulation-doped structure*. *Appl. Phys. Lett.* **68**, 418 (American Institute of PhysicsAIP, 1998). doi:10.1063/1.116704.

14. Wen, F., Shabani, J. & Tutuc, E. *Josephson Junction Field-Effect Transistors for Boolean Logic Cryogenic Applications*. *IEEE Trans. Electron Devices* **66**, 5367–5374 (Institute of Electrical and Electronics Engineers Inc., 2019). doi:10.1109/TED.2019.2951634.
15. Alam, S., Islam, M. M., Hossain, M. S. & Aziz, A. *Superconducting Josephson Junction FET-based Cryogenic Voltage Sense Amplifier*. *2022 Device Res. Conf.* 1–2 (IEEE, 2022). doi:10.1109/DRC55272.2022.9855654.
16. Jaklevic, R. C., Lambe, J., Silver, A. H. & Mercereau, J. E. *Quantum interference effects in Josephson tunneling*. *Phys. Rev. Lett.* (1964). doi:10.1103/PhysRevLett.12.159.
17. De Simoni, G., Paolucci, F., Solinas, P., Strambini, E. & Giazotto, F. *Metallic supercurrent field-effect transistor*. *Nat. Nanotechnol.* (2018). doi:10.1038/s41565-018-0190-3.
18. McCaughan, A. N. & Berggren, K. K. *A superconducting-nanowire three-terminal electrothermal device*. *Nano Lett.* (2014). doi:10.1021/nl502629x.
19. Costanzo, D., Zhang, H., Reddy, B. A., Berger, H. & Morpurgo, A. F. *Tunnelling spectroscopy of gate-induced superconductivity in MoS2*. *Nat. Nanotechnol. 2018 136* **13**, 483–488 (Nature Publishing Group, 2018). doi:10.1038/s41565-018-0122-2.
20. Zhang, H., Berthod, C., Berger, H., Giamarchi, T. & Morpurgo, A. F. *Band Filling and Cross Quantum Capacitance in Ion-Gated Semiconducting Transition Metal Dichalcogenide Monolayers*. *Nano Lett.* **19**, 8836–8845 (American Chemical Society, 2019). doi:10.1021/ACS.NANOLETT.9B03667/ASSET/IMAGES/MEDIUM/NL9B03667_M015.GIF.
21. Chernyshov, A., Overby, M., Liu, X., Furdyna, J. K., Lyanda-Geller, Y. & Rokhinson, L. P. *Evidence for reversible control of magnetization in a ferromagnetic material by means of spin–orbit magnetic field*. *Nat. Phys. 2009 59* **5**, 656–659 (Nature Publishing Group, 2009). doi:10.1038/nphys1362.
22. Liu, M., Nan, T., Hu, J. M., Zhao, S. S., Zhou, Z., Wang, C. Y., Jiang, Z. De, Ren, W., Ye, Z. G., Chen, L. Q. & Sun, N. X. *Electrically controlled non-volatile switching of magnetism in multiferroic heterostructures via engineered ferroelastic domain states*. *NPG Asia Mater. 2016 89* **8**, e316–e316 (Nature Publishing Group, 2016). doi:10.1038/am.2016.139.
23. Suleiman, M., Sarott, M. F., Trassin, M., Badarne, M. & Ivry, Y. *Nonvolatile voltage-tunable ferroelectric-superconducting quantum interference memory devices*. *Appl. Phys. Lett.* **119**, 112601 (American Institute of Physics Inc., 2021). doi:10.1063/5.0061160.
24. Alam, S., Islam, M. M., Hossain, M. S., Ni, K., Narayanan, V. & Aziz, A. *Cryogenic Memory Array based on Ferroelectric SQUID and Heater Cryotron*. *2022 Device Res. Conf.* 1–2 (IEEE, 2022). doi:10.1109/DRC55272.2022.9855813.
25. Ware, F., Gopalakrishnan, L., Linstadt, E., McKee, S. A., Vogelsang, T., Wright, K. L., Hampel, C. & Bronner, G. Do superconducting processors really need cryogenic memories? The case for cold DRAM. in *ACM Int. Conf. Proceeding Ser.* (2017). doi:10.1145/3132402.3132424.
26. Alam, S., Hossain, M. S. & Aziz, A. *A non-volatile cryogenic random-access memory based on the quantum anomalous Hall effect*. *Sci. Rep.* (2021). doi:10.1038/s41598-021-87056-7.
27. Holmes, D. S., Ripple, A. L. & Manheimer, M. A. *Energy-Efficient Superconducting Computing—Power Budgets and Requirements*. *IEEE Trans. Appl. Supercond.* **23**, 1701610–1701610 (Institute of Electrical and Electronics Engineers (IEEE), 2013). doi:10.1109/tasc.2013.2244634.
28. Alam, S., Hossain, M. S. & Aziz, A. *A cryogenic memory array based on superconducting memristors*. *Appl. Phys. Lett.* **119**, 082602 (AIP Publishing LLCAIP Publishing, 2021). doi:10.1063/5.0060716.
29. Resch, S., Cilasun, H. & Karpuzcu, U. R. *Cryogenic PIM: Challenges Opportunities*. *IEEE Comput. Archit. Lett.* **20**, 74–77 (Institute of Electrical and Electronics Engineers Inc., 2021). doi:10.1109/LCA.2021.3077536.
30. Yoon, I., Khan, A., Datta, S., Raychowdhury, A., Chang, M., Ni, K., Jerry, M., Gangopadhyay, S., Smith, G. H., Hamam, T., Romberg, J. & Narayanan, V. *A FerroFET-Based In-Memory Processor for Solving Distributed and Iterative Optimizations via Least-Squares Method*. *IEEE J. Explor. Solid-State Comput. Devices Circuits* **5**, 132–141 (Institute of Electrical and Electronics Engineers Inc., 2019). doi:10.1109/JXCDC.2019.2930222.
31. Alam, S., Islam, M. M., Hossain, M. S., Jaiswal, A. & Aziz, A. *CryoCiM: Cryogenic compute-in-memory based on the quantum anomalous Hall effect*. *Appl. Phys. Lett.* **120**, 144102 (AIP Publishing LLCAIP Publishing, 2022). doi:10.1063/5.0092169.
32. Crassous, A., Bernard, R., Fusil, S., Bouzehouane, K., Le Bourdais, D., Enouz-Vedrenne, S., Briatico, J., Bibes, M., Barthélémy, A. & Villegas, J. E. *Nanoscale electrostatic manipulation of magnetic flux quanta in ferroelectric/superconductor BiFeO 3/YBa 2Cu 3O 7-δ heterostructures*. *Phys. Rev. Lett.* **107**, 247002 (American Physical Society, 2011). doi:10.1103/PHYSREVLETT.107.247002/FIGURES/3/MEDIUM.

33. Crassous, A., Bernard, R., Fusil, S., Bouzehouane, K., Briatico, J., Bibes, M., Barthélémy, A. & Villegas, J. E. *BiFeO3/YBa2Cu3O7−δ heterostructures for strong ferroelectric modulation of superconductivity*. *J. Appl. Phys.* **113**, 024910 (American Institute of PhysicsAIP, 2013). doi:10.1063/1.4774248.
34. Huang, Q., Chen, Z., Cabral, M. J., Wang, F., Zhang, S., Li, F., Li, Y., Ringer, S. P., Luo, H., Mai, Y. W. & Liao, X. *Direct observation of nanoscale dynamics of ferroelectric degradation*. *Nat. Commun. 2021 121* **12**, 1–7 (Nature Publishing Group, 2021). doi:10.1038/s41467-021-22355-1.
35. Bardeen, J., Cooper, L. N. & Schrieffer, J. R. *Theory of superconductivity*. *Phys. Rev.* (1957). doi:10.1103/PhysRev.108.1175.
36. Alam, S., Jahangir, M. A. & Aziz, A. *A Compact Model for Superconductor- Insulator-Superconductor (SIS) Josephson Junctions*. *IEEE Electron Device Lett.* **41**, 1249–1252 (2020). doi:10.1109/LED.2020.3002448.
37. Ambegaokar, V. & Baratoff, A. *Tunneling between superconductors*. *Phys. Rev. Lett.* (1963). doi:10.1103/PhysRevLett.11.104.
38. Nguyen, M. H. *et al. Cryogenic Memory Architecture Integrating Spin Hall Effect based Magnetic Memory and Superconductive Cryotron Devices*. *Sci. Rep.* (2020). doi:10.1038/s41598-019-57137-9.
39. Mayer, W., Yuan, J., Wickramasinghe, K. S., Nguyen, T., Dartiailh, M. C. & Shabani, J. *Superconducting proximity effect in epitaxial Al-InAs heterostructures*. *Appl. Phys. Lett.* **114**, 103104 (AIP Publishing LLCAIP Publishing, 2019). doi:10.1063/1.5067363.
40. Rocci, M., De Simoni, G., Giazotto, F., Puglia, C., Esposti, D. D., Strambini, E., Zannier, V. & Sorba, L. *Gate-controlled suspended titanium nanobridge supercurrent transistor*. *ACS Nano* (2020). doi:10.1021/acsnano.0c05355.

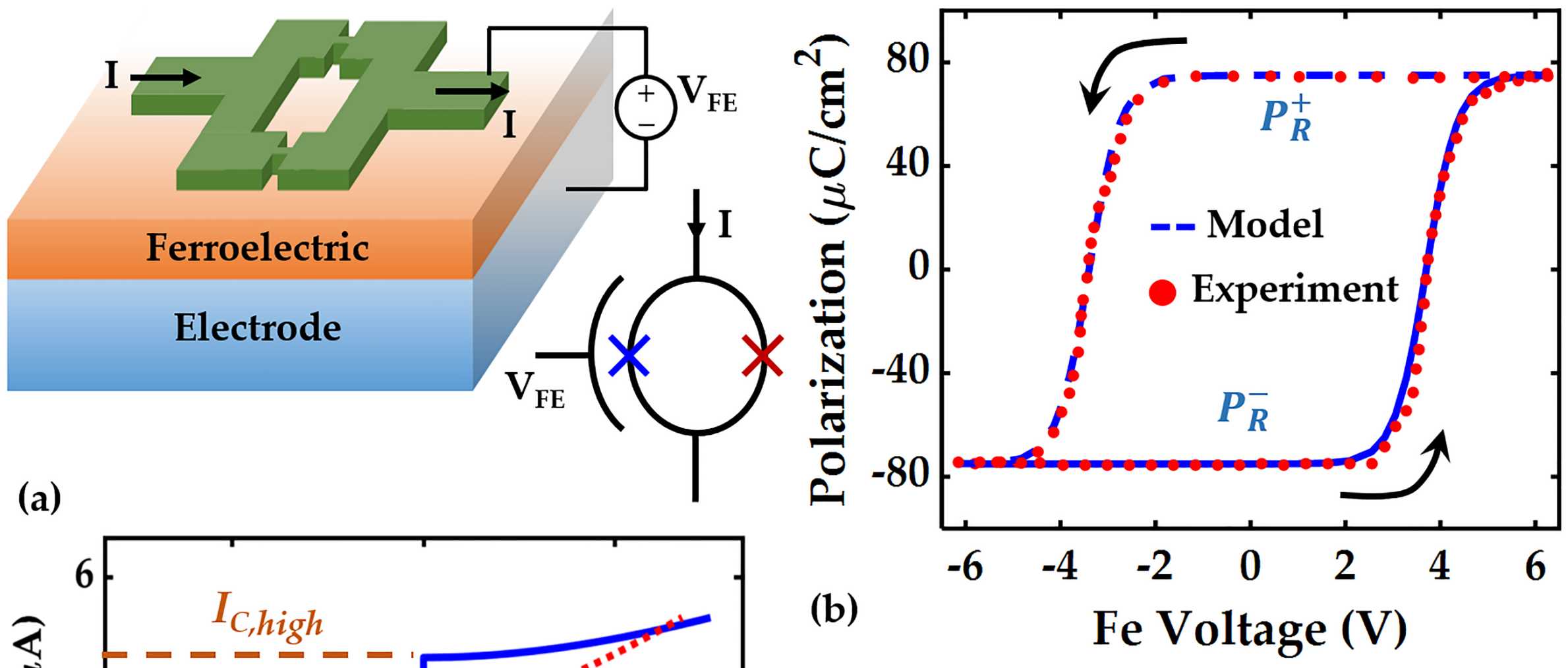

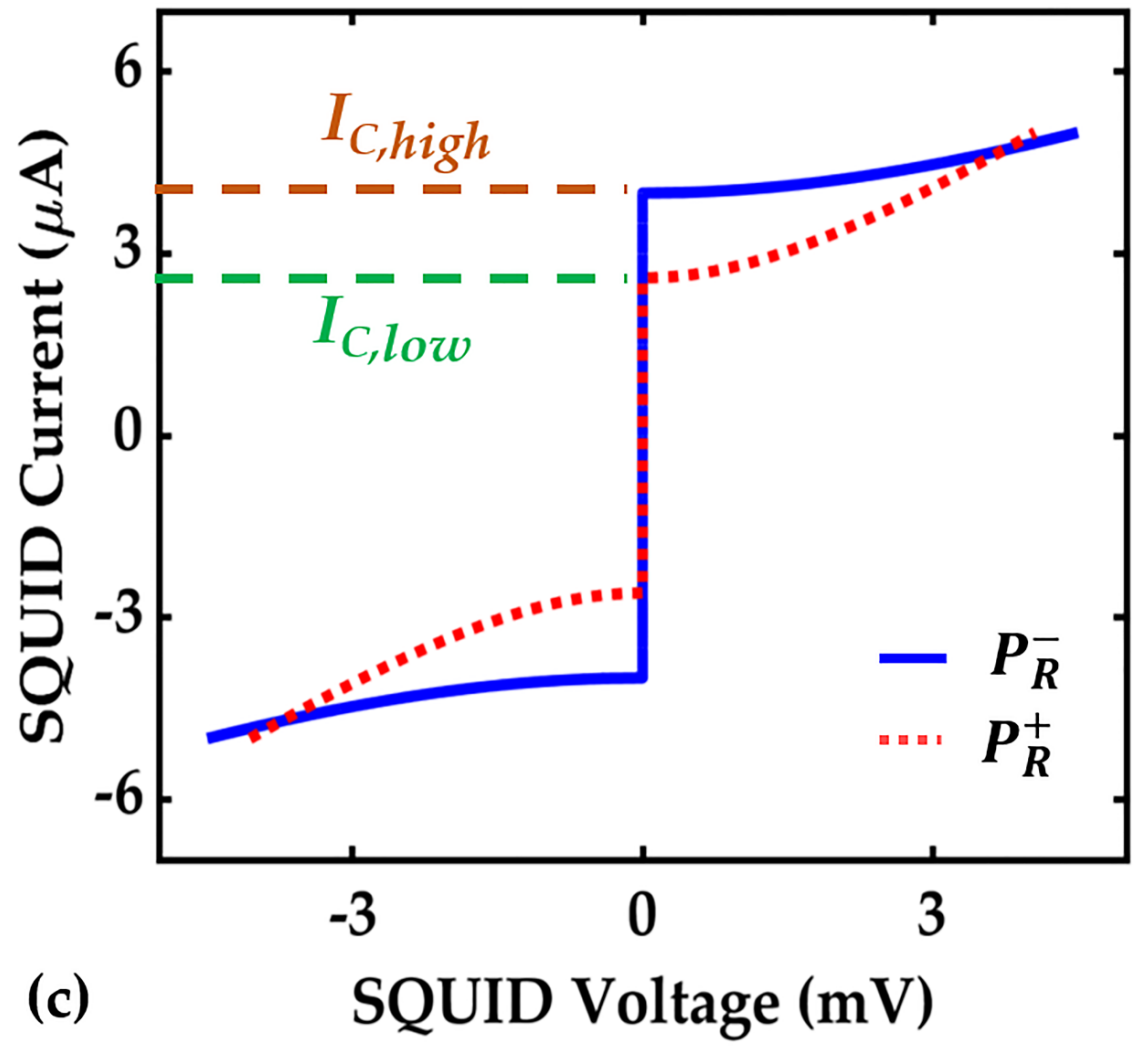

(d)

| $I_{C,low} < I < I_{C,high}$ | | | |
|---|---|---|---|
| $V_{FE}$ | $P_{FE}$ | Device State | Logic State |
| -6 V | $P_R^-$ | Superconducting | '0' |
| +6 V | $P_R^+$ | Non-superconducting | '1' |

**Fig. 1: (a)** Device structure and circuit symbol of FeSQUID. **(b)** Polarization-voltage characteristics for the PZT ferroelectric along with the validation of the developed compact model with the experimental results obtained from Ref. [23]. **(c)** *Current-voltage* characteristics of FeSQUID. Two polarization states of the ferroelectric leads to two levels of $I_C$. **(d)** Table shows the definition of logic states and corresponding device states used in the design of Boolean logic gates.

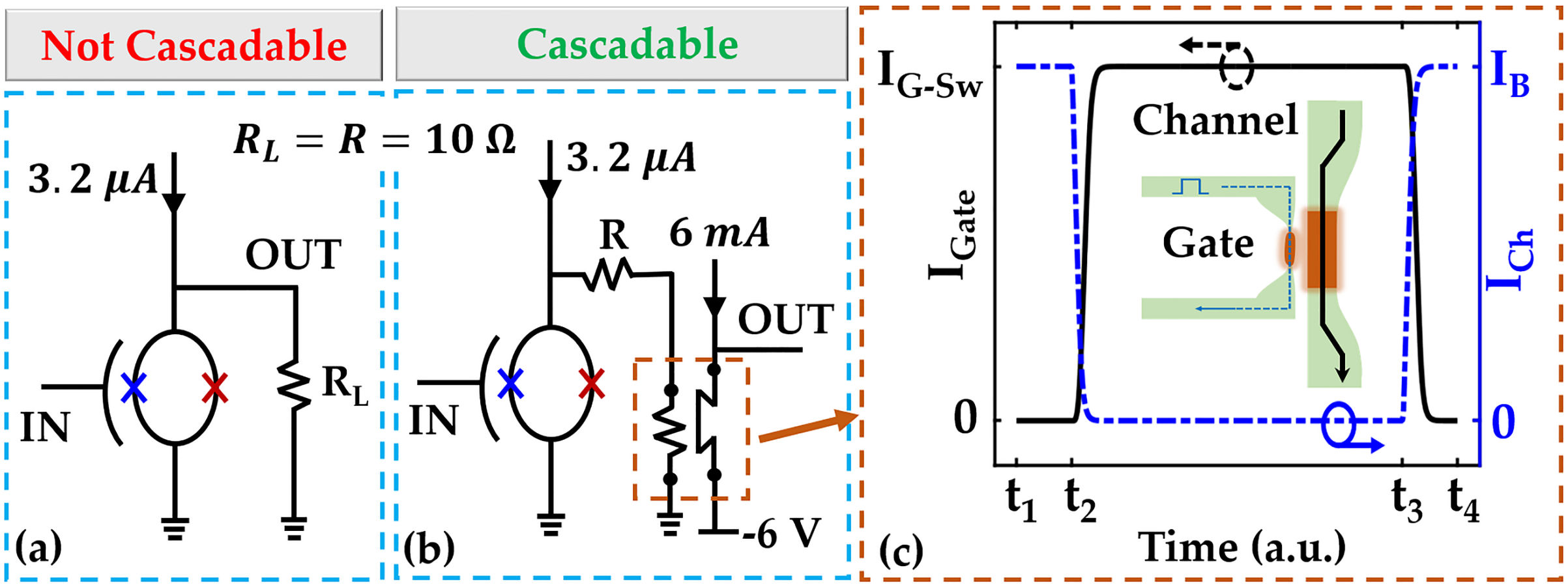

**Fig. 2:** **(a)** Schematic of a Copy gate using a FeSQUID. The output of this configuration is not sufficiently large to drive the input of the next gate and hence, this configuration is not cascadable. **(b)** Modified version of the Copy gate of (a) where a *hTron* is used to convert the output voltage to the same level of the input so that two or more logic gates can be cascaded. **(c)** *I-V* characteristics of *hTron* showing the gate current-controlled superconducting to resistive switching.

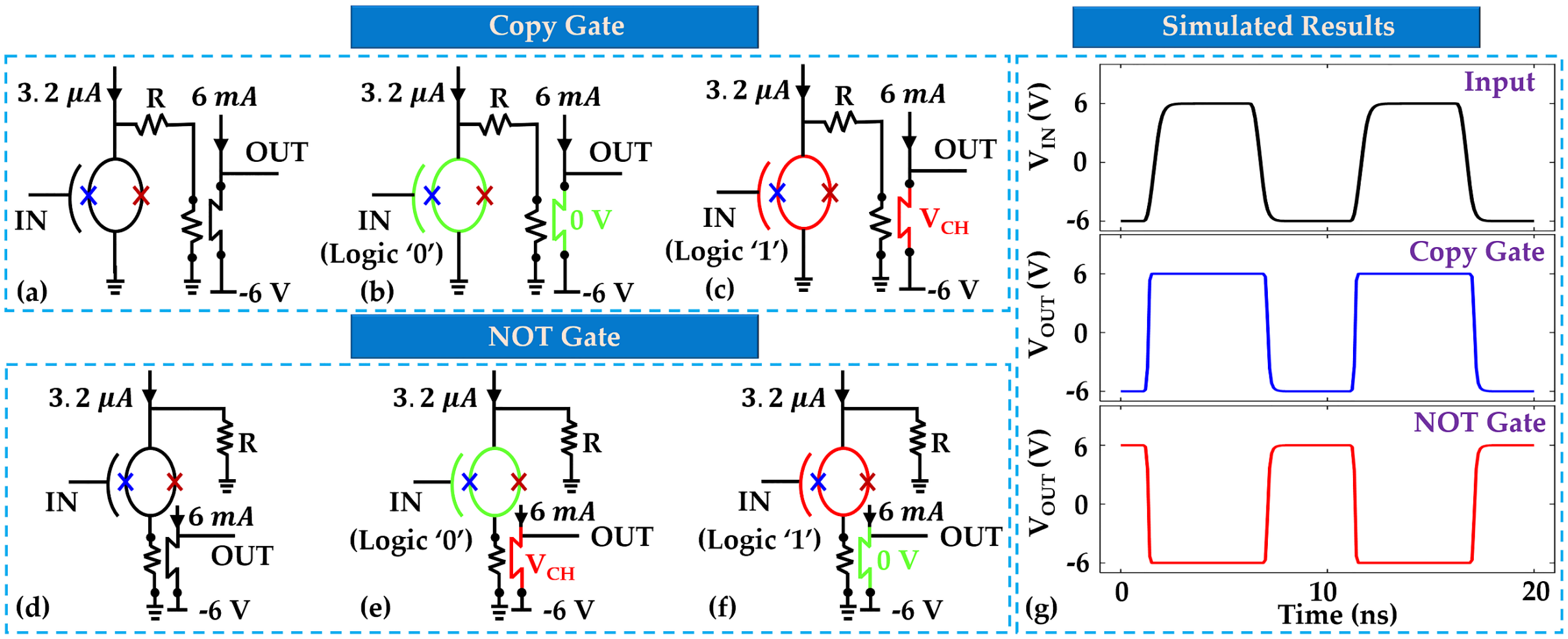

**Fig. 3:** **(a)** Schematic of FeSQUID-based Copy gate. Working principle of the Copy gate for **(b)** logic '0' and **(c)** logic '1' at the input. **(d)** Schematic of FeSQUID-based NOT gate. Working principle of the NOT gate for **(e)** logic '0' and **(f)** logic '1' at the input. Green and red colors correspond to the superconducting and resistive states, respectively. **(g)** Simulated results for the 1-input Copy and NOT gates.

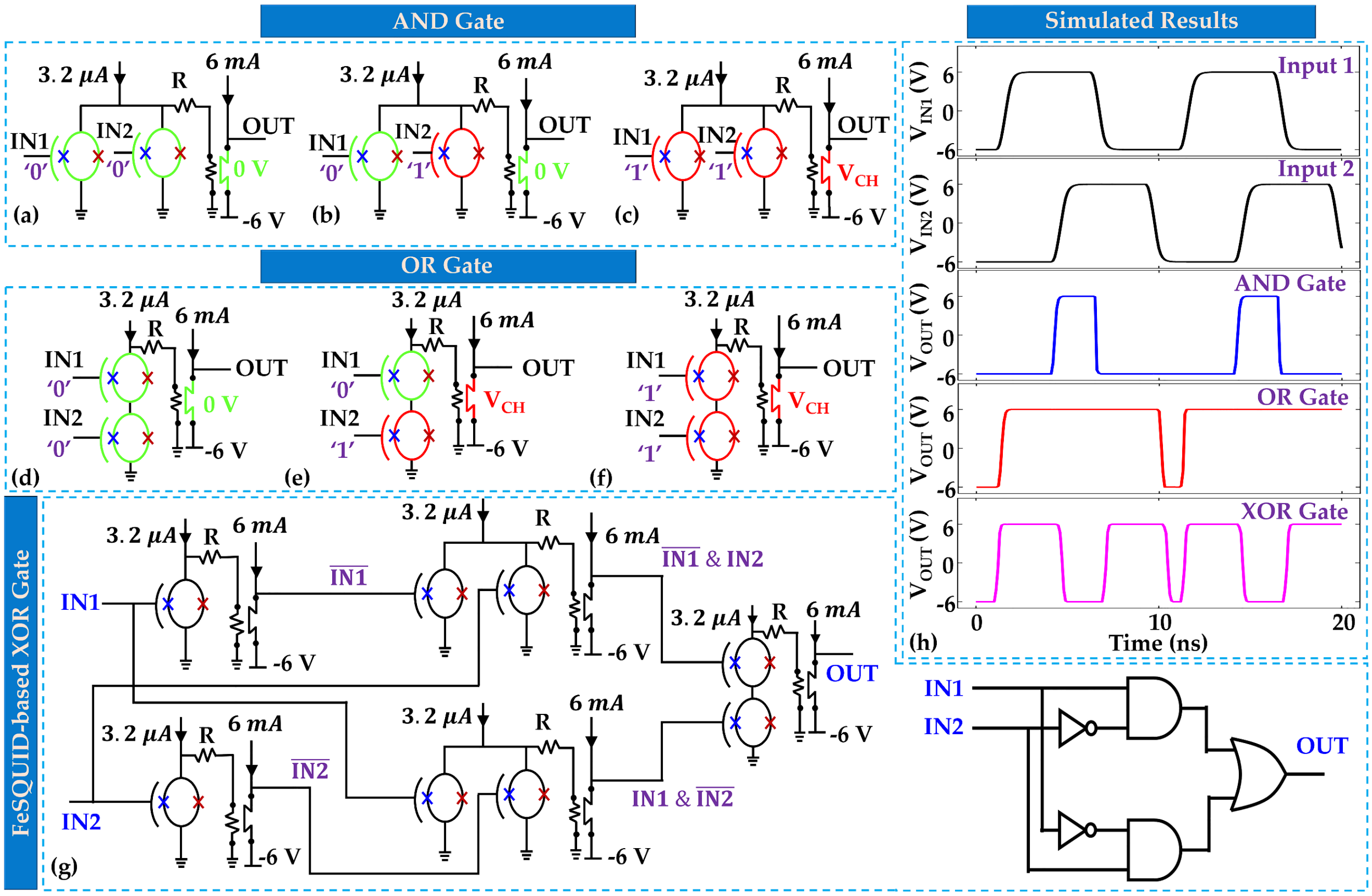

**Fig. 4:** Working principle of FeSQUID-based AND gate for inputs **(a)** '00', **(b)** '01'/'10', and **(c)** '11'. Operation of FeSQUID-based OR gate for **(d)** '00', **(e)** '01'/'10', and **(f)** '11' at the input terminals. **(g)** Schematic of XOR gate built cascading FeSQUID-based NOT, AND, and OR gates. **(h)** Simulated results for the 2-input AND, OR, and XOR gates.